\documentclass[showpacs,twocolumn]{revtex4}
\usepackage{graphicx}

\begin{document}

\title{ $\psi'$ Production and B Decay in Heavy Ion Collisions at LHC}
\author{Baoyi Chen$^1$, Yunpeng Liu$^2$, Kai Zhou$^1$ and Pengfei Zhuang$^1$}
\affiliation{$^1$Physics Department, Tsinghua University, Beijing 100084, China}
\affiliation{$^2$Cyclotron Institute, Texas A$\&$M University, College Station, TX 77843, USA}
\date{\today}

\begin{abstract}
In comparison with $J/\psi$, the excited charmonium state $\psi'$ is loosely bounded and its yield is dominantly from the B-hadron decay. Based on the transport approach, we study the double ratio of $N(\psi')/N(J/\psi)$ from A+A collisions to that from p+p collisions at LHC energy. We found that the primordial production in the initial stage and the regeneration in the hot medium are not significant for $\psi'$ production in heavy ion collisions at LHC and the double ratio in semi-central and central collisions is controlled by the B decay.
\end{abstract}
\maketitle

$J/\psi$ is a tightly bound state of charm quarks $c$ and $\bar c$. Its dissociation temperature $T_d^{J/\psi}$ in hot medium is much
higher than the critical temperature $T_c$ of the deconfinement phase transition. Therefore, the observed $J/\psi$s in
relativistic heavy ion collisions carry the information of the hot medium and have long been considered as a signature of the new state of matter, the so-called
quark-gluon plasma (QGP)~\cite{matsui}. The experimentally measured $J/\psi$ distributions, such as the nuclear modification factor~\cite{datarhic,datastar,datacms,dataalice1,dataalice2}, the elliptic flow~\cite{v2rhic,dataalice2}, and the averaged transverse momentum square~\cite{datarhic,pt2lhc} at the Relativistic Heavy Ion Collider (RHIC) and Large Hadron Collider (LHC) show significant nuclear matter effects.

In comparison with the ground state $J/\psi$, the excited state $\psi'$ has its own advantage in probing the hot medium. From the Schr\"odinger equation with lattice simulated heavy quark potential~\cite{hqp1,hqp2} at finite temperature, the dissociation temperature $T_d^{\psi\prime}$ is much lower than $T_d^{J/\psi}$, it is only a little bit above $T_c$. When the temperature of the fireball created in heavy ion collisions is above but close to $T_c$, $J/\psi$ is almost not affected by the medium, while both the primordially produced $\psi'$s in the initial stage and the regenerated~\cite{rapp,thews,pbm} $\psi'$s in the medium are crucially eaten up by the fireball. This means that $\psi'$ is more sensitive to the QGP formation. The other advantage of $\psi'$ production is that in contrast to $J/\psi$ there is no significant feed down from heavier charmonium states to $\psi'$, the production mechanism becomes more clean.

\begin{figure}[!hbt]
\centering
\includegraphics[width=0.48\textwidth]{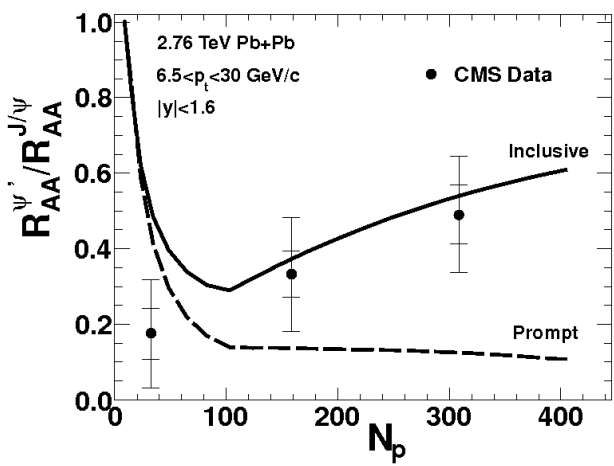}
\caption{ The double ratio $R_{AA}^{\psi'}/R_{AA}^{J/\psi}$ as a
function of number of participants $N_p$. The data are from the CMS
collaboration~\cite{cms1}, and the solid and dashed lines are
respectively the transport model calculations with and without
considering the B decay contribution. } \label{fig1}
\end{figure}

\begin{figure}[!hbt]
\centering
\includegraphics[width=0.48\textwidth]{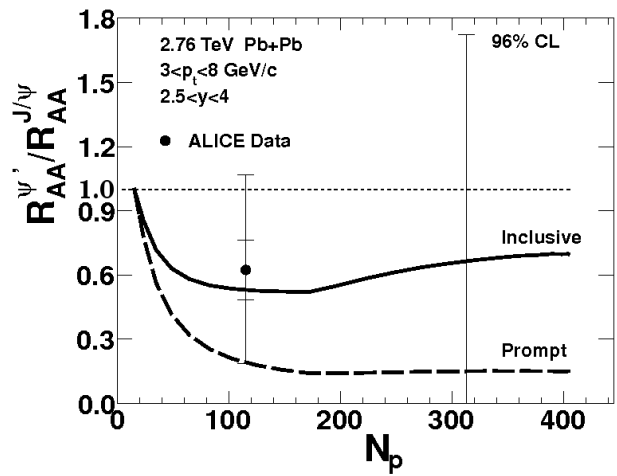}
\caption{ The double ratio $R_{AA}^{\psi'}/R_{AA}^{J/\psi}$ as a
function of number of participants $N_p$. The data are from the
ALICE collaboration~\cite{dataalice2}, and the solid and dashed
lines are respectively the transport model calculations with and
without considering the B decay contribution. } \label{fig2}
\end{figure}

\begin{figure}[!hbt]
\centering
\includegraphics[width=0.48\textwidth]{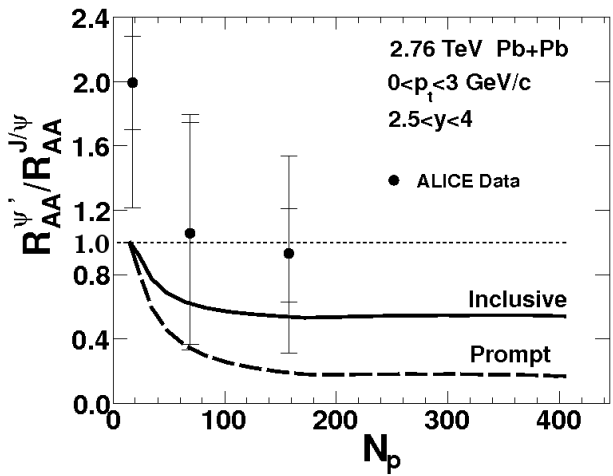}
\caption{ The double ratio $R_{AA}^{\psi'}/R_{AA}^{J/\psi}$ as a
function of number of participants $N_p$. The data are from the
ALICE collaboration~\cite{dataalice2}, and the solid and dashed
lines are respectively the transport model calculations with and
without considering the B decay contribution. } \label{fig3}
\end{figure}

\begin{figure}[!hbt]
\centering
\includegraphics[width=0.48\textwidth]{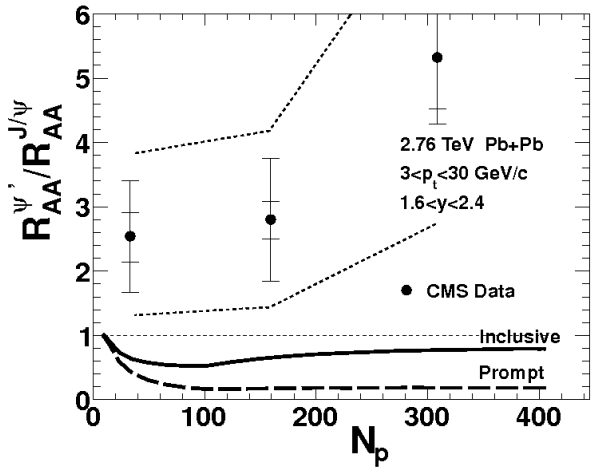}
\caption{ The double ratio $R_{AA}^{\psi'}/R_{AA}^{J/\psi}$ as a
function of number of participants $N_p$. The data are from the CMS
collaboration~\cite{cms1}, the two thin dotted lines represent
uncertainty from p+p measurement, and the solid and dashed lines are
respectively the transport model calculations with and without
considering the B decay contribution. } \label{fig4}
\end{figure}

Recently, the ALICE~\cite{dataalice2} and CMS~\cite{cms1} collaborations at LHC measured the double ratio for inclusive $\psi'$s and $J/\psi$s,
\begin{equation}
{N_{AA}^{\psi\prime}/N_{AA}^{J/\psi}\over N_{pp}^{\psi\prime}/N_{pp}^{J/\psi}}={R_{AA}^{\psi\prime}\over R_{AA}^{J/\psi}},
\label{dr}
\end{equation}
where $N_{AA}^\Psi$ and $N_{pp}^\Psi$ $(\Psi=J/\psi,\ \psi')$ are inclusive $\Psi$ numbers in Pb+Pb and p+p collisions at colliding energy $\sqrt {s_{NN}}=2.76$ TeV, and $R_{AA}^\Psi=N_{AA}^\Psi/(n_{coll}N_{pp}^\Psi)$ is the inclusive nuclear modification factor with $n_{coll}$ being the number of binary collisions determined by nuclear geometry. Beyond the expectation of the Debye screening picture~\cite{matsui} from which the double ratio would drop down with increasing centrality and almost disappear in central collisions, the observed double ratio significantly deviates from zero and does not show the decreasing trend, see Figs.\ref{fig1}-\ref{fig4} for different transverse momentum and rapidity regions. In this work, we focus on the double ratio and try to explain its surprising characteristics.

The inclusive $\Psi$ number consists of a prompt and a non-prompt part, $N_{pp}^\Psi=\overline N_{pp}^\Psi+N_{pp}^{B\to\Psi}$ and $N_{AA}^\Psi=\overline N_{AA}^\Psi+N_{AA}^{B\to\Psi}$ with $N_{AA}^{B\to\Psi}=n_{coll}N_{pp}^{B\to\Psi}Q$, where $\overline N_{pp}^\Psi$ and $\overline N_{AA}^\Psi$ are the prompt $\Psi$ numbers in p+p and A+A collisions. The non-prompt contribution $N_{pp}^{B\to\Psi}$ comes from the B-hadron decay in p+p collisions, and $Q$ is the quench factor for bottom quarks due to their energy loss in the QGP which leads to a $\Psi$ number shift from high momentum to low momentum. We have neglected here the cold nuclear matter effects~\cite{vogt1} on the bottom quark distribution. Taking into account the B decay contribution, the inclusive nuclear modification factor can be separated into two parts,
\begin{equation}
R_{AA}^\Psi={\overline R_{AA}^\Psi+r_B^\Psi Q\over 1+r_B^\Psi},
\label{nmf}
\end{equation}
where $\overline R_{AA}^\Psi=\overline N_{AA}^\Psi/(n_{coll}\overline N_{pp}^\Psi)$ stands for the prompt nuclear modification factor, and $r_B^\Psi=N_{pp}^{B\to\Psi}/\overline N_{pp}^\Psi$ is the ratio of non-prompt to prompt $\Psi$ number in p+p collisions.

The prompt charmonium motion in the QGP can be controlled by a detailed transport approach~\cite{hufner1,zhu1} which well describes the $J/\psi$ nuclear modification factor $\overline R_{AA}^{J/\psi}$~\cite{yanli,liu1,liu3}, elliptic flow~\cite{liu2} and averaged transverse momentum~\cite{zhou1,zhou2} at RHIC and LHC. To extract medium information from the charmonium motion, both the hot medium and charmonia created in high energy nuclear collisions must be treated dynamically. The charmonium phase space distribution $\overline f_\Psi({\bf p},{\bf x},t)$ is governed by a Boltzmann-type transport equation~\cite{yanli},
\begin{equation}
{\partial \overline f_\Psi\over \partial t} +{\bf v}\cdot{\bf \nabla}\overline f_\Psi=-\alpha_\Psi \overline f_\Psi +\beta_\Psi,
\label{trans}
\end{equation}
where both the initially produced and regenerated charmonia are
taken into account through the initial distribution
$\overline f_\Psi$ at the medium formation time $\tau_0$ and the gain term
$\beta_\Psi$, and all the charmonia suffer from the gluon dissociation~\cite{peskin1,chen1}
that is described by the loss term $\alpha_\Psi$. The suppression and regeneration are related to each other through detailed balance. The cold nuclear matter
effects which happen before the hot medium formation, including nuclear absorption~\cite{hufner2}, Cronin effect~\cite{cronin1} and shadowing effect~\cite{vogt1}, can be contained
in the initial distribution. From the $D$ meson spectra measured at RHIC~\cite{drhic} and LHC~\cite{dlhc}, charm quarks seem thermalized in the hot medium. Therefore, we choose thermalized gluon and charm quark distributions in the loss and gain terms. The production cross section of charm quarks in p+p collisions which controls the degree of regeneration is taken as $d\sigma_{pp}^{c\bar c}/dy=0.38$ mb~\cite{ccbarfor1,ccbarfor2} in forward rapidity and $0.62$ mb~\cite{ccbarmid} in central rapidity at LHC energy. The interaction between the charmonia and the medium depends on the local temperature $T$ and velocity $u_\mu$ of the medium which are characterized by the hydrodynamic equations~\cite{heinz1,hirano1} and the equation of state~\cite{heinz1,state1}.

The transport equation (\ref{trans}) can be analytically solved. With the distribution function $\overline f_\Psi$ at the hadronization time of the QGP, one can calculate any charmonium spectrum and compare with experimental data (we have neglected here the charmonium interaction with the hadron gas). The calculated double ratio $\overline R_{AA}^{\psi\prime}/\overline R_{AA}^{J/\psi}$ for prompt charmonia as a function of the number of participant nucleons $N_p$ at LHC is shown in Figs. \ref{fig1}-\ref{fig4} as dashed lines. The fact of lower dissociation temperature $T_d^{\psi\prime}$ means that $\psi'$ suffers more suppression in the
medium in comparison with $J/\psi$, and therefore the prompt double ratio drops down monotonously. Since the fireball formed in a heavy ion collision is inhomogeneous, the temperature in the region close to the surface is lower than $T_d^{\psi\prime}$ even for a central collision. Therefore, the produced $\psi'$s are not totally dissociated and the double ratio in central and semi-central collisions is not zero but a small value around 0.1-0.2. Since the regenerated charmonia through recombination of thermalized charm quarks carry low momentum, the contribution to the yield can be neglected in the high transverse momentum region $6.5<p_t<30$ GeV/c (Fig.\ref{fig1}) but becomes important in the low $p_t$ region $0<p_t<3$ GeV/c (Fig.\ref{fig3}).

The big difference between the dashed lines for prompt charmonia and the data shown in Figs.\ref{fig1}-\ref{fig4} may come from the non-prompt contribution. The non-prompt part of the double ratio is controlled by two parameters, the ratio $r_B^\Psi$ of non-prompt to prompt $\Psi$ and the bottom quark
quench factor $Q$. They can be extracted from p+p and p+A data or estimated from physics analysis. Since B-hadrons are easier to decay into the excited charmonium states than into the ground state, and the prompt $\psi'$s are easier to be dissociated in comparison with the prompt $J/\psi$s, the B decay contribution for $\psi'$ is much more important than that for $J/\psi$. Considering the fact that the charmonia from B decay carry high momentum, the ratio $r_B^\Psi$ should increase with transverse momentum $p_t$. For $J/\psi$, the p+p data~\cite{rbjpsi1} at $\sqrt {s_{NN}}=2.76$ TeV show a linear increase of the non-prompt fraction with transverse momentum, $N_{pp}^{B\to J/\psi}/N_{pp}^{J/\psi}=0.04+0.023 p_t$/(GeV/c), and the result is not sensitive to the rapidity and the colliding energy. From the simple relation $N_{pp}^{B\to J/\psi}/N_{pp}^{J/\psi}=r_B^{J/\psi}/(1+r_B^{J/\psi})$, we can easily extract the non-prompt to prompt ratio $r_B^{J/\psi}$. Lack of non-prompt $\psi'$ data at $\sqrt {s_{NN}}=2.76$ TeV, we extract the non-prompt to prompt ratio $r_B^{\psi\prime}$ from the p+p data at $\sqrt {s_{NN}}=1.96$ TeV~\cite{rbpsiprime}, assuming that it is not sensitive to the colliding energy. For the transverse momentum and rapidity regions corresponding to the ALICE and CMS data shown in Figs.\ref{fig1}-\ref{fig4}, we take the average ratios $(r_B^{J/\psi},\ r_B^{\psi\prime})$=(0.27, 0.38) for $6.5<p_t<30$ GeV/c and $|y|<1.6$, (0.16, 0.26) for $3<p_t<8$ GeV/c and $2.5<y<4$, (0.1, 0.2) for $0<p_t<3$ GeV/c and $2.5<y<4$, and (0.17, 0.28) for $3<p_t<30$ GeV/c and $1.6<y<2.4$.

The bottom quark quench factor $Q$ reflects the hot medium effect in A+A collisions, it shifts the charmonia from high momentum to low momentum. It is in general larger than unit at low $p_t$ and smaller than unit at high $p_t$. For the high $p_t$ region $6.5<p_t<30$ GeV/c at rapidity $|y|<1.6$ shown in Fig.\ref{fig1}, we extract $Q=0.35$ directly from the experimental data~\cite{cms1}. For the other regions in Figs.\ref{fig2}-\ref{fig4}, there are no available data and the model predictions~\cite{qmodel2,qmodel3,kim1} are still with sizeable uncertainties. We take $Q=1$ for the intermediate $p_t$ regions in Figs.\ref{fig2} and \ref{fig4} and 1.4 for the low $p_t$ region in Fig.\ref{fig3}, simply considering the above physics analysis. These values are within the region of the model calculations~\cite{qmodel2,qmodel3,kim1}.

The calculated inclusive double ratio $R_{AA}^{\psi\prime}/R_{AA}^{J/\psi}$ as a function of centrality is shown in Figs.\ref{fig1}-\ref{fig4} as solid lines. Note that from the experimental data~\cite{cms1}, the quench factor $Q$ is almost centrality independent in the region of $100\lesssim N_p$. Therefore, the centrality dependence of the double ratio is controlled by the prompt part $\overline R_{AA}^\Psi$. In the low $N_p$ region where the prompt contribution is remarkable, the inclusive double ratio $R_{AA}^{\psi\prime}/R_{AA}^{J/\psi}$ decreases like the prompt one $\overline R_{AA}^{\psi\prime}/\overline R_{AA}^{J/\psi}$. However, in the intermediate and large $N_p$ region where most of the prompt $\psi'$s are dissociated in the hot medium, the centrality dependence of the inclusive double ratio is characterized only by the $J/\psi$ nuclear modification factor $\overline R_{AA}^{J/\psi}$. With the continuous decrease of $\overline R_{AA}^{J/\psi}$ in this region, see the experimental data~\cite{dataalice1,dataalice2,cms1} and our model calculation~\cite{liu1}, the double ratio $R_{AA}^{\psi\prime}/R_{AA}^{J/\psi}$ goes up. The kink at $N_p\sim 100$ in mid rapidity (Figs.\ref{fig1} and \ref{fig4}) and $N_p\sim 180$ at forward rapidity (Fig.\ref{fig2}) corresponds to the $J/\psi$ dissociation temperature $T_d^{J/\psi}$. The disappearance of the kink in Fig.\ref{fig3} is due to the strong regeneration in the low $p_t$ region. The model calculations including the B decay contribution agree reasonably well with the ALICE and CMS data shown in Figs.\ref{fig1}-\ref{fig3}.
\begin{figure}[!hbt]
\centering
\includegraphics[width=0.48\textwidth]{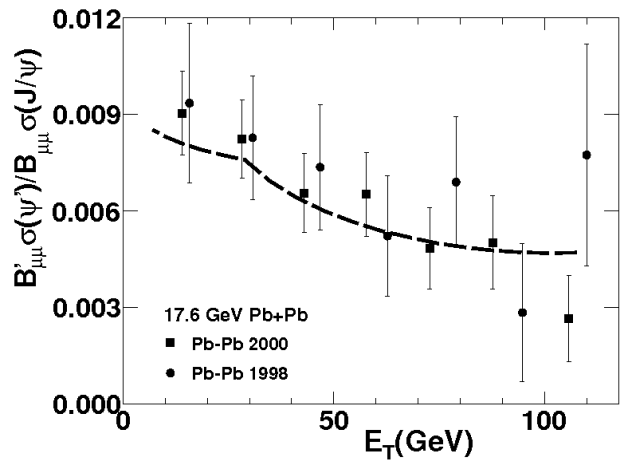}
\caption{ The yield ratio $\psi'/J/\psi$ as a function of transverse energy. The data are from the NA50 collaboration~\cite{na50}, and the line is the transport model calculation without considering charmonium regeneration and B decay. }
\label{fig5}
\end{figure}
\begin{figure}[!hbt]
\centering
\includegraphics[width=0.48\textwidth]{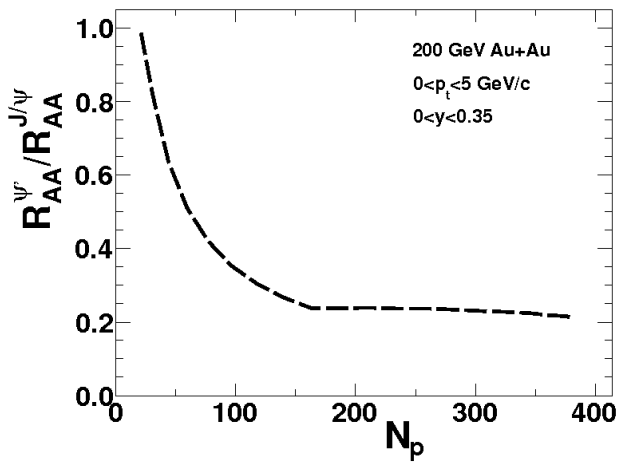}
\caption{The double ratio $R_{AA}^{\psi'}/R_{AA}^{J/\psi}$ as a function of number of participants $N_p$. In the calculation the charmonium regeneration is included but the B decay is neglected. }
\label{fig6}
\end{figure}

While the transverse momentum and rapidity windows in Fig.\ref{fig4} are close to those in Figs.\ref{fig1}-\ref{fig3}, the inclusive double ratio behaves very differently. It is less than or around unit in Figs.\ref{fig1}-\ref{fig3}, but much larger than unit in Fig.\ref{fig4} in the whole centrality region. To see the possibility to reach such a large double ratio in our model, we consider the maximum B decay contribution and the minimal hot medium effect for $\psi'$. We put all the $\psi'$s from the B decay into the region $3<p_t<30$ GeV/c of Fig.\ref{fig4} by adjusting the quench factor $Q=2.3$ and take the minimal $\psi'$ suppression by letting $\overline R_{AA}^{\psi\prime}=\overline R_{AA}^{J/\psi}$. The obtained double ratio is still much lower than the current data. For instance, $R_{AA}^{\psi\prime}/R_{AA}^{J/\psi}=1.3$ at $N_p=300$ can even not reach the lower limit of the data.

We calculated also the yield ratio $\psi'/J/\psi$ for Pb+Pb collisions at SPS energy $\sqrt {s_{NN}}=17.3$ GeV. At such low energy, there are no B decay and no charmonium regeneration in the medium. The only hot nuclear matter effect on the charmonium production is the suppression. Fig.\ref{fig5} shows our model calculation based on the transport approach~\cite{zhu1} and the NA50 data~\cite{na50}. Different from the collisions at RHIC and LHC energy where the nuclear absorption in the initial stage can be neglected in comparison with the strong hot nuclear matter effect, the absorption at SPS energy is important~\cite{hufner1}. In our calculation we take the nuclear absorption cross section as $\sigma_{abs}=4.3,\ 4.3,\ 7.9$ mb~\cite{na50} for $J/\psi$, $\chi_c$ and $\psi'$, respectively. The transport approach describes the double ratio very well.

The good agreement of the model calculation with the data at SPS energy is not enough to support the background of the calculation at LHC, since the charmonium regeneration inside the QGP may play an important role in the double ratio. To this end we calculate the charmonium distributions at RHIC energy where the charmonium regeneration becomes important~\cite{rapp,thews,pbm,yanli} but the B decay contribution is still small. From our previous calculations on $J/\psi$, the model describes very well the nuclear modification factors $\overline R_{AA}(N_p)$ and $\overline R_{AA}(p_t)$~\cite{zhu1,liu3}, the elliptic flow~\cite{liu2,zhou2}, and the averaged transverse momentum~\cite{zhou1}. The calculated double ratio at RHIC energy is shown in Fig.\ref{fig6}. At the moment there are no $\psi'$ data at RHIC. From the comparison of the calculation at LHC including B decay with the calculations at SPS and RHIC without B decay, a significant difference is the increasing double ratio in semi-central and central collisions at LHC due to the B decay contribution.

\begin{figure}[!hbt]
\centering
\includegraphics[width=0.48\textwidth]{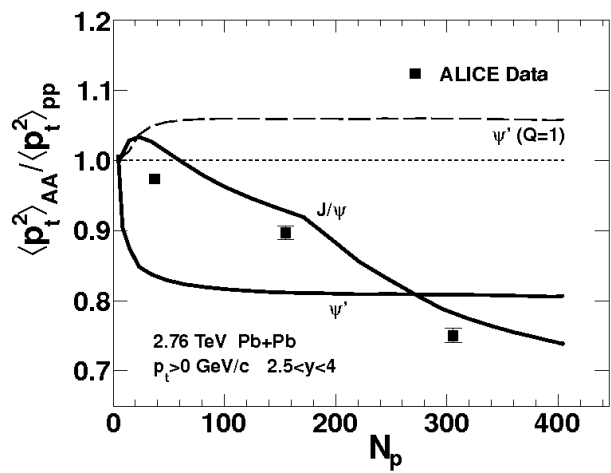}
\caption{The ratio $r_{AA}=\langle p_t^2\rangle_{AA}/\langle
p_t^2\rangle_{pp}$ as a function of number of participants $N_p$ for
$J/\psi$ and $\psi'$. The $J/\psi$ data are from \cite{pt2lhc}, and
the solid and dashed lines are the transport model calculations with
and without considering bottom quark quench in hot medium. }
\label{fig7}
\end{figure}

The particle transverse motion is more sensitive to the dynamics of the collision system. From the recent theoretical~\cite{liu1,liu3} and experimental~\cite{datarhic,pt2lhc} studies, the $J/\psi$ transverse momentum distributions at SPS, RHIC and LHC behave very differently. To further see the difference between the $J/\psi$ and $\psi'$ production mechanisms, we calculate the averaged transverse momentum square
\begin{equation}
\label{pt2}
\langle p_t^2\rangle_{AA}^\Psi= \langle\overline{p_t^2}\rangle_{AA}^\Psi {\overline N_{AA}^\Psi\over N_{AA}^\Psi}+\langle p_t^2\rangle_{AA}^{B\to\Psi}{N_{AA}^{B\to\Psi}\over N_{AA}^\Psi},
\end{equation}
where $\langle\overline{p_t^2}\rangle_{AA}^\Psi$ and $\langle p_t^2\rangle_{AA}^{B\to\Psi}$ are the averaged transverse momentum square of prompt and non-prompt $\Psi$s. The ratio
\begin{equation}
\label{ratio}
r_{AA}^\Psi={\langle p_t^2\rangle_{AA}^\Psi\over \langle p_t^2\rangle_{pp}^\Psi}
\end{equation}
for $\Psi=J/\psi$ and $\psi'$ and the comparison with the $J/\psi$
data~\cite{pt2lhc} in Pb+Pb collisions at forward rapidity and
colliding energy $\sqrt{s_{NN}}=2.76$ TeV are shown in
Fig.\ref{fig7}. For $J/\psi$, the B decay contribution in A+A
collisions can be neglected and the ratio is controlled by the
prompt part, $r_{AA}^{J/\psi}\simeq \langle \overline {p_t^2}
\rangle_{AA}^{J/\psi}/\langle p_t^2\rangle_{pp}^{J/\psi}$. It
increases in the beginning due to the Cronin effect~\cite{cronin1}
and then drops down all the way due to the more and more important
regeneration which happens in the low $p_t$ region. The model
calculation agrees reasonably well with the data. For $\psi'$, the
prompt part in A+A collisions is eaten up by the medium and the
ratio is characterized by the B decay, $r_{AA}^{\psi'}\simeq \langle
p_t^2 \rangle_{AA}^{B\to\psi'}/\langle p_t^2\rangle_{pp}^{\psi'}$.
In this case the nuclear matter effect comes purely from the quench
factor $Q$. Since $Q$ does not depend on the centrality for
semi-central and central collisions~\cite{cms1}, the ratio becomes
almost a constant in a wide region of $N_p$. From $\langle
p_t^2\rangle_{AA}^{B\to\psi'}(Q) \le \langle
p_t^2\rangle_{AA}^{B\to\psi'}(Q=1)=\langle
p_t^2\rangle_{pp}^{B\to\psi'}$ due to the quench effect, we have
$r_{AA}^{\psi'}\le \langle p_t^2\rangle_{pp}^{B\to\psi'}/\langle
p_t^2\rangle_{pp}^{\psi'}>1$. The up limit of the ratio at $Q=1$
means a full perturbative QCD calculation for the B decay process in
A+A collisions.

In summary, we calculated the double ratio $(N_{AA}^{\psi\prime}/N_{AA}^{J/\psi})/(N_{pp}^{\psi\prime}/N_{pp}^{J/\psi})$ for inclusive charmonia in heavy ion collisions at LHC energy. After experiencing the hot medium created in the early stage of the collisions, a part of the promptly produced $J/\psi$s still survive and dominate the final state $J/\psi$ distributions, but most of the prompt $\psi'$s are dissociated in the hot medium, and the finally observed $\psi'$s are mainly from the B-hadron decay. Therefore, the double ratio in semi-central and central heavy ion collisions is controlled by the B decay. Our transport approach calculations agree reasonably well with the LHC data in most transverse momentum and rapidity regions, but fail to explain the data in the region of $3<p_t<30$ GeV/c and $1.6<y<2.4$. The big difference between the theory and the data and the large uncertainty in the data need further theoretical study and precise experimental measurement.
\appendix {\bf Acknowledgement}: We thank Dr. Nu Xu for helpful discussions. The work is supported by the NSFC under grant No. 11079024 and the MOST under grant No. 2013CB922000.


\begin{thebibliography}{20}
\bibitem{matsui} T.Matsui and H.Satz, Phys. Lett. {\bf B178}, 416(1986).
\bibitem{datarhic} PHENIX Collaboration, A.Adare {\it et al.}, Phys. Rev. Lett. {\bf 98}, 232301(2007).
\bibitem{datastar} STAR Collaboration, B.I.Abelev {\it et al.}, Phys.Rev. {\bf C80}, 041902(R)(2009)
\bibitem{datacms} C.Silvestre, [for the CMS Collaboration], J. Phys. G: Nucl. Part. Phys. {\bf 38}, 124033(2011).
\bibitem{dataalice1} I.Arsene, [for the ALICE Collaboration], Nucl. Phys. {\bf A904}, 623c(2013).
\bibitem{dataalice2} E.Scomparin, [for the ALICE Collaboration], arXiv:1211.1623[nucl-ex].
\bibitem{v2rhic} Z.Tang, [for the STAR Collaboration], J. Phys. G: Nucl. Part. Phys. {\bf 38}, 124107(2011).
\bibitem{pt2lhc} E.Scomparin, [for the ALICE Collaboration], Talk at Quark Matter Conference 2012.
\bibitem{hqp1} D.Chen, R.C.Brower, J.W.Negele, and E.V.Shuryak, Nucl. Phys. Proc. Suppl. {\bf 73}, 512(1999).
\bibitem{hqp2} E.V.Shuryak and I.Zahed, Phys. Rev. {\bf D70}, 054507(2004).
\bibitem{rapp} L.Grandchamp and R.Rapp, Nucl. Phys. {\bf A715}, 545c(2003).
\bibitem{thews} R.L.Thews, Eur. Phys. J. {\bf A29}, 15(2006).
\bibitem{pbm} S.J.Lindenbaum and R.S.Longacre, Phys. Rev. {\bf C78}, 054904(2008).
\bibitem{cms1} D.Moon, [for the CMS Collaboration], arXiv:1209.1084v2[hep-ex].
\bibitem{vogt1} R.Vogt, Phys. Rev. {\bf C81}, 044903(2010).
\bibitem{hufner1} J.Hufner and P.Zhuang, Phys. Lett. {\bf B559}, 193(2003).
\bibitem{zhu1} X.Zhu, P.Zhuang and N.Xu, Phys. Lett. {\bf B607}, 107(2005).
\bibitem{yanli} L.Yan, P.Zhuang and N.Xu, Phys. Rev. Lett. {\bf 97}, 232301(2006).
\bibitem{liu1} Y.Liu, Z.Qu, N.Xu and P.Zhuang, Phys. Lett. {\bf B678}, 72(2009).
\bibitem{liu3} Y.Liu, B.Chen, N.Xu and P.Zhuang, Phys. Lett. {\bf B697}, 32(2011).
\bibitem{liu2} Y.Liu, N. Xu and P.Zhuang, Nucl. Phys. {\bf A834}, 317c(2010).
\bibitem{zhou1} K.Zhou, N.Xu and P.Zhuang, Nucl. Phys. {\bf A834}, 249c(2010).
\bibitem{zhou2} K.Zhou, N.Xu, Z.Xu, P.Zhuang, paper in preparation.
\bibitem{peskin1} M.E.Peskin, Nucl. Phys. {\bf B156}, 365(1979).
\bibitem{chen1} B.Chen, K.Zhou and P.Zhuang, Phys. Rev. {\bf C86}, 034906(2012).
\bibitem{hufner2} C.Gerschel and J.Hufner, Annu. Rev. Nucl. Part. Sci. {\bf 49}, 255(1999).
\bibitem{cronin1} S.Gavin and M.Gyulassy, Phys. Lett. {\bf B214}, 241(1988).
\bibitem{drhic} H.Zhang, [for the STAR Collaboration], J. Phys. G: Nucl. Part. Phys. {\bf 32}, S29(2006).
\bibitem{dlhc} A.Dainese, [for the ALICE Collaboration], J. Phys. G: Nucl. Part. Phys. {\bf 38}, 124032(2011).
\bibitem{ccbarfor1} M.Cacciari, M.Greco, P.Nason, JHEP, {\bf 05}, 007(1998).
\bibitem{ccbarfor2} M.Cacciari, S.Frixione, P.Nason, JHEP, {\bf 03}, 006(2001).
\bibitem{ccbarmid} ALICE Collaboration, B.Abelev {\it et al.}, JHEP {\bf 07}, 191(2012).
\bibitem{heinz1} P.F.Kolb, J.Sollfrank and U.Heinz, Phys. Rev. {\bf C62}, 054909(2000).
\bibitem{hirano1} T.Hirano, Phys. Rev. {\bf C65}, 011901(2001); T.Hirano {\it et al.}, Phys. Lett. {\bf B636}, 299(2006).
\bibitem{state1} J.Sollfrank {\it et al.}, Phys. Rev. {\bf C55}, 392(1997).
\bibitem{rbjpsi1} CMS Collaboration, S.Chatrchyan {\it et al.}, JHEP {\bf 05}, 063(2012).
\bibitem{rbpsiprime} CDF Collaboration, T.Aaltonen {\it et al.}, Phys. Rev. {\bf D80}, 031103(2009).
\bibitem{qmodel2} W.A.Horowitz, and M.Gyulassy, J. Phys. G: Nucl. Part. Phys. {\bf 35}, 104152(2008).
\bibitem{qmodel3} M.He, R.J.Fries and R.Rapp, Phys. Rev. {\bf C86}, 014903(2012).
\bibitem{kim1} H.Kim, Talk at The 4th Asian Triangle Heavy Ion Conference.
\bibitem{na50} H.Santos, [for the NA50 Collaboration], J. Phys. G: Nucl. Part. Phys.  {\bf 30}, S1175(2004).
\end{thebibliography}
\end{document}